\documentclass[times]{article}

\usepackage{graphicx}
\usepackage{amssymb}
\graphicspath{{./Figures/}}

\usepackage{lineno}

\usepackage[usenames, dvipsnames]{color} 
\usepackage{amsmath}          
\usepackage[utf8x]{inputenc}  
\usepackage{graphicx}         
\usepackage{amssymb,amsfonts}
\usepackage{fixmath}
\usepackage{multirow}
\usepackage{pgfplots}
\usepackage{bbold}
\usepackage[ruled,vlined]{algorithm2e}
\usepackage{algorithmicx,algpseudocode}
\usepackage{tikz}
\usepackage[labelsep=quad,indention=10pt]{caption}
\usepackage{bm}
\usepackage{tabu}
\usepackage{float}
\usepackage{subfig}
\usepackage{booktabs}
\usepackage{siunitx}

\newcommand{\bh}{\bm{h}}

\newcommand{\nMC} {\ensuremath{\texttt{n}_{\texttt{MC}}}}

\newcommand{\nd} {\ensuremath{\texttt{n}_{\texttt{d}}}}
\newcommand{\ncon} {\ensuremath{\texttt{n}_{\texttt{con}}}}
\newcommand{\ns} {\ensuremath{\texttt{n}_{\texttt{s}}}}

\newcommand\RR{\leavevmode\hbox{$\rm I\!R$}}

\newcommand{\MRAText}[1]{{\color{black} #1}}

\begin{document}

\title{Adaptive surrogates of crashworthiness models for multi-purpose engineering analyses accounting for uncertainty}

\author{Marc Rocas(1,2), Alberto Garc\'ia-Gonz\'alez(1), Xabier Larrayoz(2),\\ Pedro D\' iez(1,3)\\ \\
$1$- Laboratori de C\`alcul Num\`eric, E.T.S. de Ingenier\'ia de Caminos,\\ Universitat Polit\`ecnica de Catalunya -- BarcelonaTech\\
$2$- SEAT, Martorell, Barcelona\\
$3$- The International Centre for Numerical\\ Methods in Engineering, CIMNE, Barcelona}

\maketitle

\begin{abstract}
Uncertainty Quantification (UQ) is a booming discipline for complex computational models based on the analysis of robustness, reliability and credibility. UQ analysis for nonlinear crash models with high dimensional outputs presents important challenges.  In crashworthiness, nonlinear structural behaviours with multiple hidden modes require expensive models (18 hours for a single run).  Surrogate models (metamodels) allow substituting the full order model, introducing a response surface for a reduced training set of numerical experiments.  Moreover, uncertain input and large number of degrees of freedom result in high dimensional problems, which derives to a bottle neck that blocks the computational efficiency of the metamodels. Kernel Principal Component Analysis (kPCA) is a multidimensionality reduction technique for non-linear problems, with the advantage of capturing the most relevant information from the response and improving the efficiency of the metamodel. Aiming to compute the minimum number of samples with the full order model. The proposed methodology is tested with a practical industrial problem that arises from the automotive industry.\\


\end{abstract}


\section{Introduction}
\label{S:1}
Uncertainty Quantification (UQ) plays an important role in crash analysis, aiming to obtain robust models that provide sensitivity analysis and probabilistic information. In crashworthiness, having an accurate knowledge of the model allows to reduce the number of prototypes and, in consequence, their costs.
Stochastic modeling for UQ has been used in the crash industry for decision making support \cite{rocas2020nonintrusive,wang2018crashworthiness}.
The process to propagate the variability from the input to the output requires thousands of model evaluations. For the automotive industry, where each single model consumes at least 18 hours, it is essential to reduce the number of simulations to a small training set of experiments. Beyond the computational cost, crashworthiness presents important challenges in metamodeling high-dimensional problems for outputs with large degrees of freedom presenting nonlinear behaviours. Reference \cite{2102.07673} presents a comparative study for crash simulations combining dimensionality reduction and surrogate modelling through a probability density function analysis. Other surrogate models based in similar strategies are recently presented in \cite{sancarlos2021pgdbased}.
Also, Min Li proposed a sensitivity analysis methodology combining PCA and Kriging for models with high-dimensional outputs. The study was applied for the San Francisco coastal protection \cite{li2020efficient}. Also J.B Nagel, J. Rieckermann and B. Sudret proposed a sensitivity analysis methodology applying PCA for the reduction of the outputs and Polynomial Chaos for surrogate modeling for an urban drainage model \cite{nagel2017uncertainty}. All of these studies present strategies for dealing with dimensionality reduction and surrogate modeling for UQ. However, large number of simulations for the training set are needed, and in consequence high computational cost for the field of crashworthiness.
 
In this work, an adaptive (or levelled) methodology that combines dimensionality reduction and surrogate modeling for nonlinear complex models is proposed. Specifically, kernel Principal Component Analysis (kPCA) is used to reduce the high-dimensional outcomes to low number of components and Ordinary Kriging (OK) to metamodel between the stochastic input space and the reduced space from kPCA. \MRAText{Quite apart from the terms of standard UQ, the methodology allows obtaining complementary multi-purpose information of the model. Different features of the model are obtained in this context with a negligible additional computational effort: structural modes associated with output data, sensitivity analysis (influence from perturbation of input parameters in the results), statistical assessment of various quantities of interest...} The method provides an efficient and robust tool for decision making with the minimum evaluations of the full order model but guaranteeing precision. This adaptive strategy allows to evaluate only the necessary samples for the training set to optimize the computational cost. In order to demonstrate the performance of the proposed approach, a benchmark crash problem is studied. The literature contains different works in crashworthiness UQ field \cite{2102.07673, rocas2020nonintrusive,wang2018crashworthiness}, where different UQ approaches are shown implementing techniques as Monte Carlo, Polynomial Chaos, Quasi Monte Carlo, dimensionality reduction and surrogate modelling.

The paper is structured as follows: In Section \ref{S:2} a benchmark crash problem is presented. In Section \ref{S:3} is provided the description of the adaptive methodology divided in main steps. In Section \ref{S:4} it is presented the results of the benchmark problem for a vademecum of 3000 simulations and the results of the proposed methodology. \MRAText{Finally,  Sections \ref{S:5} and \ref{S:6} close the manuscript with a discussion and some conclusions.}

\section{Benchmark crash problem}
\label{S:2}

In this section, a benchmark problem is presented to validate the feasibility of the proposed UQ methodology.

\begin{figure}[H]
\centering\includegraphics[width=0.9\linewidth]{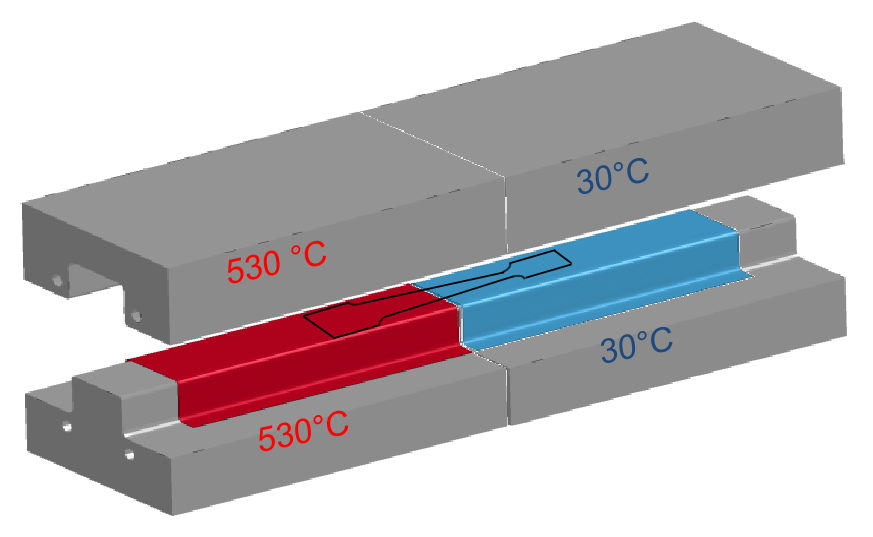}
\caption{Tailored tempering process. Tailored press with two temperatures. The Right press with $40^\circ C$ and the left press with $530^\circ C$.}
\label{fig:press}
\end{figure}

\begin{figure}[H]
\centering\includegraphics[width=0.95\linewidth]{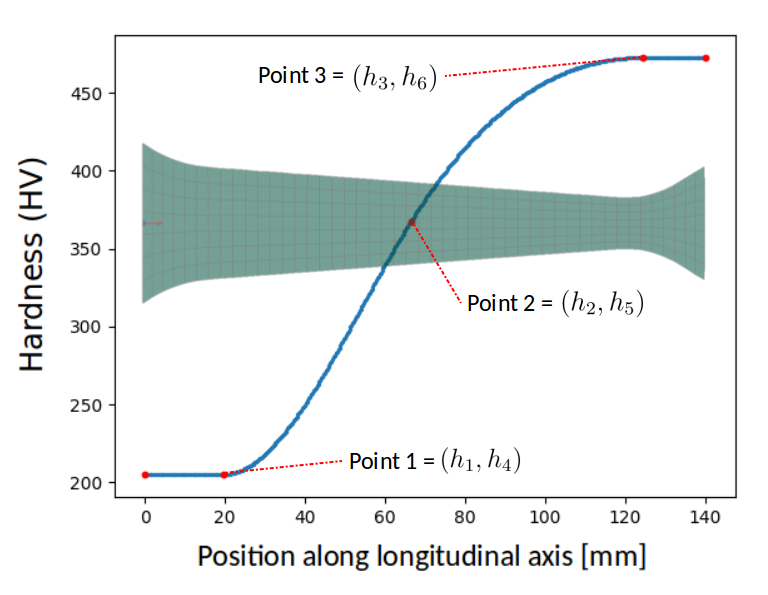}
\caption{Hardness curve for the sheet piece through the manufacture process of heated and cooled press halves.}
\label{fig:hardness_curve}
\end{figure}

In the field of crashworthiness, the B-pillar is a part of the structure of a car that plays an important role in passenger safety. The manufacture process is one of the keys to achieve a successful design. \MRAText{Initially, the B-pillars were designed by assembling monolithic parts with different strengths capabilities. The idea was to produce a substructure with variable strength depending on the external loads. Recently, this structure design has been improved by reducing the number of pieces and weight by introducing a tailored templed B-pillar with a variable hardness profile, that is variable mechanical properties.
The \emph{tailored tempering} manufacturing process results in a progressive hardness profile, as illustrated in Fig. \ref{fig:press}.}  An austenitized sheet piece with a thickness of $1.5$ mm is introduced to a tailored press. This press is divided in two halves. One half with a temperature of $40^\circ C$ and the other half with $530^\circ C$. After a holding time of 20 seconds the piece is extracted and cooled down to room temperature. As a result, the piece has a progressive hardness profile. In Fig. \ref{fig:hardness_curve} is illustrated the hardness curve. \MRAText{The problem of the tailored tempering process is to ensure certain reproducibility within a series production process.  Random perturbations of the mechanical properties are  inevitable in serial production.
Thus, the resulting mechanical properties are affected by important uncertainties, to be modelled with material parameters of aleatoric nature.
Controlling this stochastic process with robustness is a challenge for the industry, and represents a cumbersome task.} Therefore, computational modelling can be a hard task due to its random behaviour.

For this research, the model is developed with explicit formulation in VPS/Pamcrash \cite{Pamcrash}. Solving the equation of transient dynamics
\begin{equation}
\mathbf{M}  \mathbf{\ddot U} +\mathbf{C}\mathbf{\dot U}+\mathbf{K}\mathbf{U}=\mathbf{F}_{ext}(t),
\label{eq:transient}
\end{equation}

where $\mathbf{M}$ (mass matrix), $\mathbf{C}$ (damping matrix), $\mathbf{K}$ (Stiffness matrix), $\mathbf{F}_{ext}$ (external force) and $t$ (time).

\begin{figure}[H]
\centering\includegraphics[width=1\linewidth]{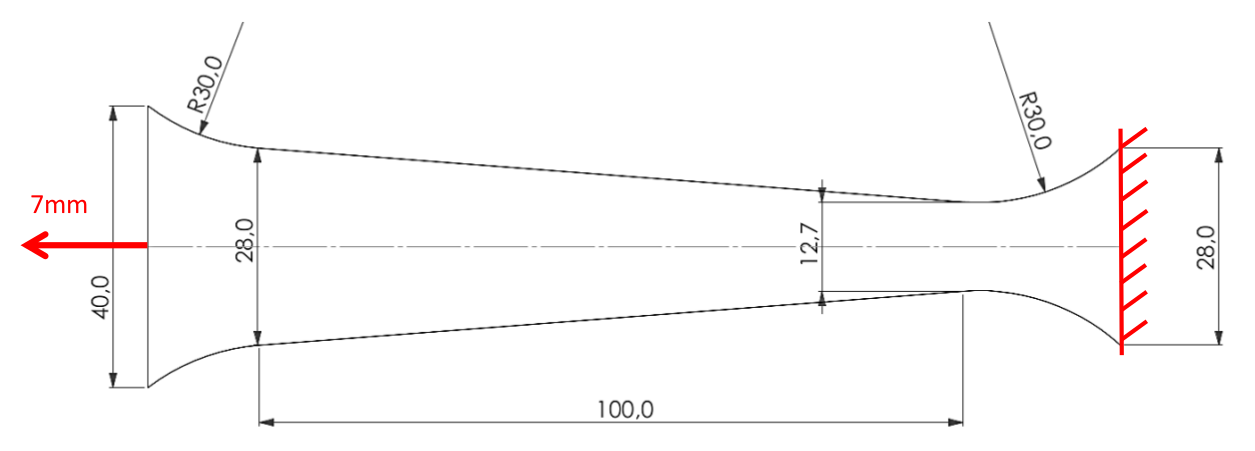}
\caption{Geometry of the benchmark model.}
\label{fig:croquis}
\end{figure}

\MRAText{A simplified model is adopted here to demonstrate the strengths and capabilities of the proposed algorithms and data driven strategies. It models a tapered tensile specimen. With respect to a realistic B-pillar profile, this benchmark has much lower computational requirements while containing the essential features of the problem, allowing to account for the same conceptual difficulties and reproduce all the pertinent mechanisms.} The geometry of the model is illustrated in Fig. \ref{fig:croquis}. The structure is fixed in the right side. In the left side a uniform displacement of $7$ mm in $40$ ms (uniaxial load) is prescribed. The benchmark is modelled using the Belytschko-Tsay shell element with one integration point in the plane. The model has a total of 329 quadratic shell elements of $1.5$ mm (thickness) and 384 nodes. A fracture model with a no element elimination configuration is implemented to guaranty the same number of elements for each simulation. For the time discretization it is used a time step of $0.2$ ms.

The model is characterized with a Young Modulus of $200$ [GPa] and a Poisson ratio of $0.3$. Due to the manufacturing process explained above, the variability of the problem comes from the hardness curve of the material. To characterize the random behaviour of the curve, the position of the three points in Fig. \ref{fig:hardness_curve} are taken as the uncertain input of the problem. Namely, Point 1=$(h_1,h_4)$, point 2=$(h_2,h_5)$ and point 3=$(h_3,h_6)$. The six stochastic variables are collected in a vector of inputs $\bh=[h_1,h_2,h_3,h_4,h_5,h_6]^T$. The random inputs are assumed to be uncorrelated with Gaussian distributions $h_i\thicksim \mathcal{N}(\mu_i,\sigma_i),i=1,2,...,6$. All the other parameters in the model are considered deterministic.  In Table \ref{tab:table1} the mean and standard deviation (StD) for each variable is described.

\begin{table}[h!]
\caption{Uncertainty variables.}
\begin{center}
\begin{tabular}{l|c|r} 
\textbf{Inputs} & \textbf{Mean} & \textbf{StD}\\
$h_i$ & $\mu_i$ & $\sigma_i$ \\
\hline
$h_1$ & 20 & $5.5\%$\\
$h_2$ & 70 & $5.5\%$\\
$h_3$ & 120 & $3\%$\\
$h_4$ & 212 & $5.5\%$\\
$h_5$ & 360 & $2.5\%$\\
$h_6$ & 460 & $3\%$\\
\end{tabular}
\label{tab:table1}
\end{center}
\end{table}

The output of the solution $\mathbf{U}$ of eq. \ref{eq:transient} is characterized as a QoI vector $\mathbf{x}$ of dimension $d=329$, corresponding to the values of the plastic strain in 329 elements. In practise, sampling the parametric input values and computing the full order model results in collecting different matrices of $\mathbf{x} \in \RR^d$.

\section{Adaptive UQ methodology}
\label{S:3}

\subsection{Dimensionality reduction, surrogate modeling and UQ}

A novel UQ methodology for high dimensional outputs in the field of crashworthiness is proposed in \cite{2102.07673}. The methodology combines Dimensionality Reduction (DR) and Surrogate Modeling (SM). The approach requires a specific number of evaluation of the high order model for the UQ methodology. Then, a DR technique is proposed to reduce the output data to a reduced space with lower dimension and SM to establish a relationship between the input space and the reduced space. Once the metamodel is achieved, a standard Monte Carlo analysis is carried out to perform an UQ study with a negligible computational cost. 

The main disadvantage of this approach is how to define the number of samples for the training set  (to guarantee enough information for an UQ analysis). In industry, the size of the data set is defined based on computational resources, objectives, and the model. In terms of efficiency, this is an inappropriate approach to deal with this kind of problems, since the computational cost of the full order model is high, where each evaluation consume high resources and time. In this paper it is presented an UQ methodology to evaluate only the necessary samples for the training set without losing precision. The novel approach is based in an adaptive strategy combining DR+SM.

\subsection{Adaptive UQ methodology}

\begin{figure}[H]
\centering\includegraphics[width=1\linewidth]{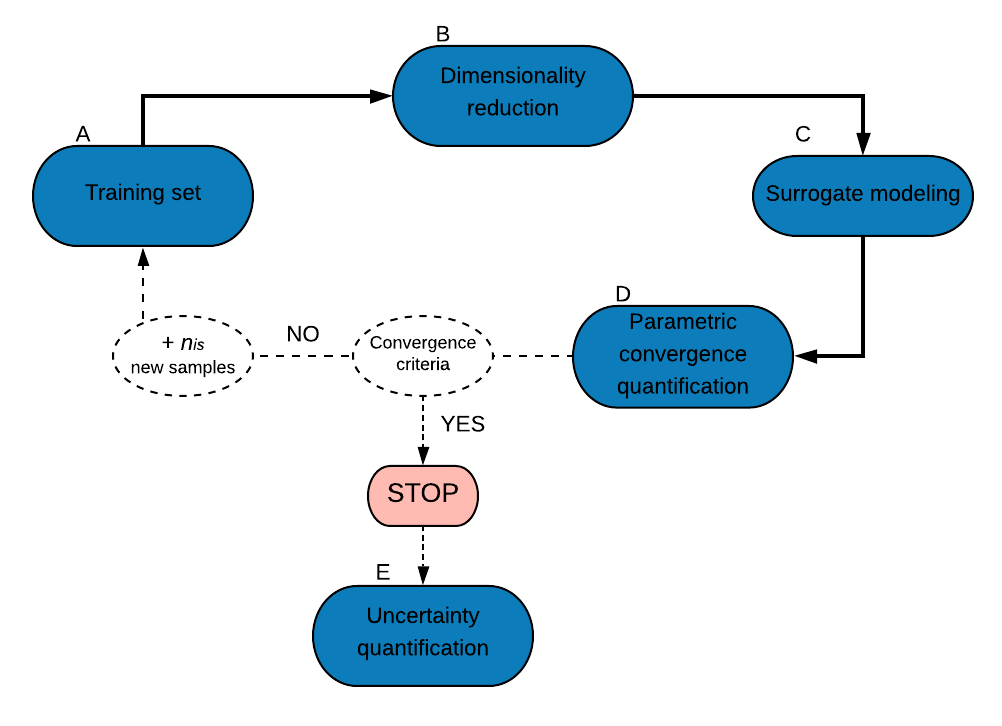}
\caption{Flowchart of the adaptive UQ methodology}
\label{fig:general_flowchart}
\end{figure}

To guarantee a robust design with a small number of evaluation of expensive models becomes a real challenge. The proposed methodology is developed to deal with the complex issue to quantify the uncertainty for crash problems, with the aim of minimizing computational cost, while preserving precision with an adaptive approach. In this section, the main steps of the proposed strategy are described. Fig. \ref{fig:general_flowchart} presents the flowchart, and in the following subsections each step is explained in detail.
A general overview of the main steps in the flowchart follows:

\begin{itemize}
\item The first step, called \textit{ A-Training set}, relies on the identification and characterization of the stochastic inputs to evaluate the expensive model in a set of training points. Then, the Quantity of Interest (QoI) of each simulation is stored in the output matrix $\mathbf{X}$. 

\item The second step, \textit{B-Dimensionality reduction} is based on the kPCA dimensionality reduction technique applied to nonlinear data set. However, other techniques can be implemented. This step is intended to reduce the dimension of the output matrix $\mathbf{X}$. The reduced space of principal components allows to detect hidden structural modes and also avoids jeopardizing the metamodel approach. 

\item The third step, \textit{C-Surrogate modeling }, corresponds to the development of a response surface between the reduced space, from kPCA, and the stochastic space of inputs. The metamodel allows to substitute the full order model to evaluate any new point in the reduced space and to map it backwards in the original space \cite{garcia2020kernel}. 

\item The fourth step, \textit{D-Parametric convergence quantification}, it is evaluated the metamodel with new Monte Carlo samples as a substitute of the expensive model. Sensitivity analysis of the input parameters (Sobol Indices) and clustering are performed in the enriched reduced space of kPCA. \MRAText{These indices are used in a stopping criterion to check the convergence and stability of the method.}

\item In the fifth step, \textit{E-Uncertainty quantification}, the enriched space of kPCA is mapped backwards to perform uncertainty quantification of the input space. Statistical measures of the input space (QoI histograms, means, variances, standard deviations and free new simulations) are analyzed with negligible computational cost.
\end{itemize}

The first four steps (A, B, C, D) are implemented in an adaptive scheme with different levels of sampling sizes. The variance of the percentage of the clusters and the sensitivity indices of the input parameters are compared for each level aiming to analyze the stability of the problem outcome. If the stopping criterion is fulfilled, the method stops. If not, it goes to the next level with a new sampling size $\ns$, for enriching the training set. This levelled approach is implemented until stability is achieved. The details of the adaptive stopping criteria are explained in Section \ref{stopping_criteria}.

\subsection{Training set}
\begin{figure}[H]
\centering\includegraphics[width=1\linewidth]{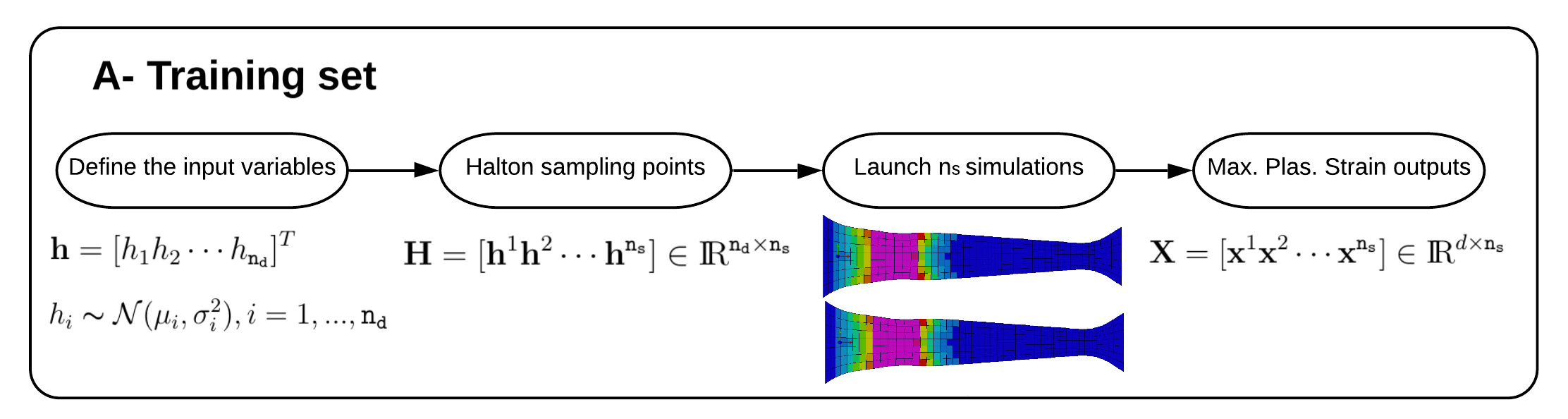}
\caption{Flowchart of the training model.}
\label{fig:training_block}
\end{figure}

Let us consider a set of random variables describing the input parameters of the model by $\bh=[h_1 h_2 \cdots h_{\nd}]^T$. A Halton sampling technique \cite{wong1997sampling} is used to select $\ns$ points (to build a discrete training set in the input space). The advantage of Halton sequence with respect other techniques as Monte Carlo, Hammersley sequences, among others, is the  nested samples property for each resampling level size. The sampling points of the input space are stored in the matrix $\mathbf{H}=[\bh^1 \bh^2 \cdots \bh^{\ns}]\in \RR^{\nd \times \ns}$. Each input vector $\bh^i,i=1,2,...,\ns$ requires a single run in VPS/pamcrash. The idea is to evaluate $\ns$ simulations of the expensive model to store the output responses in $\mathbf{X}=[\mathbf{x}^1 \mathbf{x}^2 \cdots\mathbf{x}^{\ns}]\in \RR^{d \times \ns}$ as a training set. In the benchmark problem, each $\mathbf{x}^i$
collects the maximum plastic strain of all the elements of the model in the last time step. Fig. \ref{fig:training_block} illustrates the flowchart to sample the parametric space and obtain the training set.

\subsection{kPCA dimensionality reduction} \label{DR}

Analyze the training set and find the principal components allows to reduce de complexity of the problem. The dimensionality reduction technique kPCA is proposed for the reduction of the output matrix $\mathbf{X}=[\mathbf{x}^1 \mathbf{x}^2	\cdots \mathbf{x}^{\ns}]\in \RR^{d \times \ns}$. It is of utmost importance consider that in the field of crashworthiness the data is nonlinear. The standard Principal Component Analysis (PCA) captures linear behaviours, however for this research kPCA is implemented for its non-linear ability and the extremely advantageous backward mapping to recover back the full-order object in as accurately as possible. In this aspect, kPCA behaves much better than PCA in many cases.

kPCA provides a useful tool to transform the original data from high dimensional space to a low dimensional space where the main features of the input data are kept. Considering the training set matrix $\mathbf{X}=[\mathbf{x}^1 \mathbf{x}^2 \cdots\mathbf{x}^{\ns}]\in \RR^{d \times \ns}$ as the input matrix, the main objective is to find a low dimensional space, where the first $k$ principal components retain most of the information to capture the data behaviour. For this, a nonlinear mapping function $\mathbf{\Phi}(\mathbf{x})$ is needed, where in general it is unknown. However, the most used kernel functions are:

\begin{itemize}
\item Gaussian kernel: $\kappa(\mathbf{x}^i,\mathbf{x}^j)=e^{-\beta\|\mathbf{x}^i-\mathbf{x}^j\|^2}$
\item Linear kernel: $\kappa(\mathbf{x}^i,\mathbf{x}^j)=<\mathbf{x}^i,\mathbf{x}^j>$
\item Polynomial kernel: $\kappa(\mathbf{x}^i,\mathbf{x}^j)=(<\mathbf{x}^i,\mathbf{x}^j>+b)^p$
\end{itemize}

Collecting a reduced number of terms with enough pieces of information allows to reduce the number of metamodels for the feature space $\mathbf{z}^{\star}$, and in consequence, the computational cost. This combination is problem dependent and refit kPCA multiple times to compare different kernels and parameters with a optimization function is needed. Also, if the data has different behaviours and the mapping function is appropriate, then the reduced space $\mathbf{z}^{\star}$ is a sensitive measure for cluster detection (structure modes). 

In practice, this boils down to apply kPCA and determine a mapping function $\mathcal{G}(\mathbf{x})$ between the solutions $\mathbf{x}\in \RR^{d}$ and some new variable $\mathbf{z}^{\star}\in\RR^{k}$ in a much lower-dimensional space ($k\ll d$). The set of eigenvalues provides the criterion to choose the number of terms $k$ to be retained for the reduction (for the benchmark problem, $90\%$ of information must be retained), leading to, 

\begin{equation}
\mathbf{z}^{\star}=\mathcal{G}(\mathbf{x}).
\end{equation}

The mapping between $\mathbf{x}$ and $\mathbf{z}^{\star}$ is to be characterized forward and backward as $\mathbf{x}^{\star}=\mathcal{G}^{-1}(\mathbf{z}^{\star})\approx \mathbf{x}$. There exist different techniques available in the literature \cite{zheng2010penalized,wang2012kernel}. For this research, it is implemented a technique based on a minimization of the discrepancy functional (residual) \cite{garcia2020kernel}. 

The proposed weighting distance technique \cite{garcia2020kernel} associates weights to each value in the calculation based on the distance between the samples and the input sample $\mathbf{x}$. Let $d_i$ for $i=1,...,\ns$ be the squared distances $d_{i}=\Vert \mathbf{z}^{\star} - \mathbf{z}^{\star \, i}  \Vert$. Where $w_i$ are the weights defined by,

\begin{equation}
w_i=\frac{1/d_i^2}{\sum_{j=1}^{\ns}1/d_j^2}.
\end{equation}

\MRAText{Here, the inverse of the squared distances ($1/d^2$) is used to define the weights, following \cite{garcia2020kernel}. Any other decreasing function of the distance is admissible, to account for the influence of the distance in the weights. Any version of the radial-based interpolation is commonly used to construct surrogate models based on samples from a training set.}

For a $\mathbf{z}^{\star}$ point in the reduced space it corresponds a $\mathbf{x}$ point in the orginal space defined by:

\begin{equation}
\mathbf{x}\approx\sum_{i=1}^{\ns}w_i\mathbf{x}^i.
\end{equation}

\begin{figure}[H]
\centering\includegraphics[width=1\linewidth]{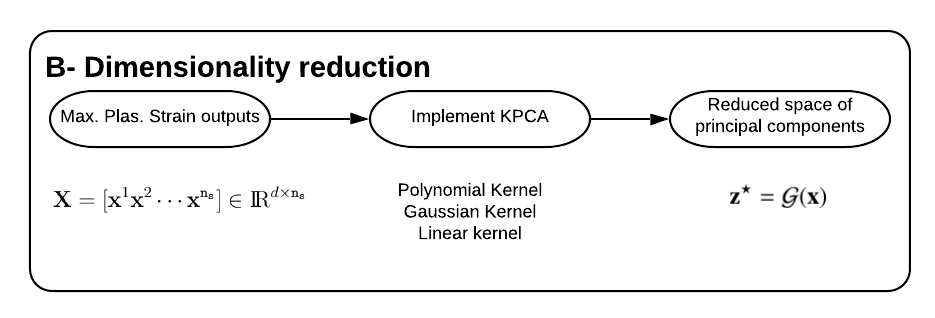}
\caption{Flowchart of the dimensionality reduction step.}
\label{fig:dim_block}
\end{figure}

This technique allows to backward any point from the reduced space $\mathbf{z}^{\star}$ to the original space. In Fig. \ref{fig:dim_block} it is illustrated the flowchart for the dimensionality reduction step.

\subsection{Surrogate modeling}

In crashworthiness, it is common to have high dimensional output matrices. In consequence, it is unaffordable to construct a surrogate model in terms of computational cost to deal for an efficient approach.
 
The main idea of this surrogate modeling step is constructing a response surface $\mathbf{z}^{\star}=F(\bh)$, from input $\bh$ to the reduced space $\mathbf{z}^{\star}$. Here, the surrogate technique is presented to establish a functional dependency among some input $\bh$ and some output function $\mathbf{y}(\bh)$ (typically, a postprocess or reduced model of $\mathbf{x}^{\star}$).

A scalar output $Y$ is considered for any of the components of $\mathbf{y}(\bh)$. For the benchmark problem, $Y$ corresponds to the first principal component of kPCA, that is $Y=[\mathbf{z}^{\star}]_{1}=[y^1y^2\cdots y^{\ns}]^T$, where $y^i, i=1,2,...,\ns $ are the points of the reduced space . 

The functional dependence $Y = F(\bh)$ is determined from the data provided by the training set, and the dimensionality reduction space. The metamodel function $F(\bh)$, approximates for any input $\bh$ the corresponding image $y^i$ in the reduced space. Then the backward mapping explained in Section \ref{DR} returns to the input sample by $\mathbf{x}^{\star}\approx \mathcal{G}^{-1}(F(\bh))$.  

For this research, Ordinary Kriging (OK) is used for metamodeling. In the literature different papers regarding kriging metamodeling \cite{oliver2014tutorial,2102.07673}. Other surrogate modelling technique can be implemented as Polynomial Chaos or Separated Response Surface \cite{rocas2020nonintrusive,2102.07673}. However, kriging shows a better performance for crash modelling \cite{2102.07673}. The purpose is to evaluate the metamodel with new $\nMC$ realizations to estimate new $\mathbf{z}^{\star}$ values to enrich the reduced space for a posteriori UQ analysis. Here it is presented a brief review of OK for the methodology.

OK is an interpolation surrogate method that determines weights for a set of sample points to obtain a prediction of a new input. The weights are based on a variogram model that has the main advantage of estimating different variances for any distance between a pair of samples. The kriging metamodel $F(\bh)$ of any point $\bh$ is defined by:
\begin{equation}
F^j (\bh)= \sum_{i=1}^{\ns} w^i_j(\bh) [\mathbf{y}^i]_j, j=1,2,...,k.
\end{equation}

\begin{figure}[H]
\centering\includegraphics[width=1\linewidth]{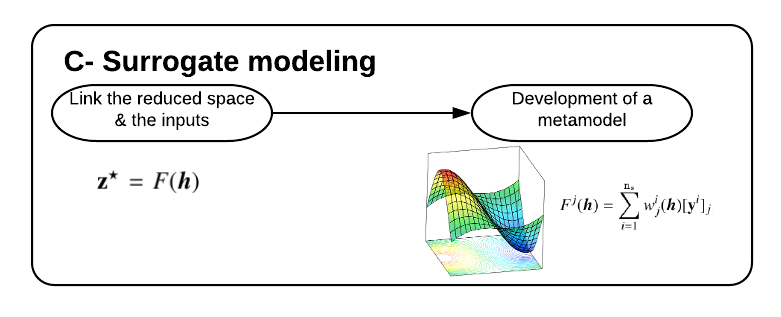}
\caption{Flowchart of the surrogate modeling.}
\label{fig:surr_block}
\end{figure}

The unknowns $w$ are the weights and $[y^i]_j$ are the scalars of the principal component $j$ of kPCA.  This means that for each dimension of $\mathbf{z}^{\star}$ is needed a particular surrogate model. Therefore, the first $k$ terms of the feature space determines the number of metamodels needed for the approach. The main condition with OK with respect other kriging approaches is that the sum of weights is equal to 1. For more theoretical details of OK see \cite{oliver2014tutorial}. 
In Fig. \ref{fig:surr_block} it is illustrated the scheme for the surrogate modeling block.

\subsection{Parametric convergence quantification}
Optimization and redesign is a common task in crashworthiness, where detecting the main structure modes and the principal parameters could provide a useful tool for the engineers. For that, once the surrogate model $F(\bh)$ is available, for each input value $\bh$, the corresponding $\mathbf{z^{\star}}$ is computed as $F(\bh)$. At this point, standard Monte Carlo is performed with $\nMC=10^5$ new random samples of $\bh$ to evaluate the surrogate model. The operations are computationally negligible with respect to the cost of the training set of the full order simulations. Therefore, sensitivity and statistical measures are easily performed. 

In order to quantify the parametric uncertainty of the problem, sensitivity analysis of the input parameters (Sobol Indices) and clustering (structure modes) are performed in the enriched reduced space $\mathbf{z}^{\star}$ of kPCA with new $\nMC = 10^5 $ samples. Here, the clustering technique K-means \cite{likas2003global} is implemented for the reduced space $\mathbf{z}^{\star} \in \RR^{k \times \ns}$ in order to detect clusters in the data. However, other cluster techniques can be implemented \cite{saxena2017review} . On the other hand, Sobol Indices are implemented to characterize the influence of the inputs to the outputs. 

The essence of Sobol Indices is based on a variance decomposition of the feature space $Y$. We can define the total variance of $Y$ as $Var_{Y}$. Therefore, $Var_{Y}$ can be decomposed into partial variance associated with the inputs $\bh$ as $Var_{Y}=\sum_{i=1}^{\nd}Var_i+\sum_{i=1}^{\nd-1}\sum_{j=i+1}^{\nd}Var_{ij}+...+Var_{1,...,\nd}$, where $Var_i$ denotes the  variance contribution of the parameter $h_i$. While the other terms make reference with high order of interaction between inputs $\bh$. Two sensitivity measures provide the Sobol Indices:

\begin{itemize}
\item Parameter influence ranking
\item Identification of negligible parameters
\end{itemize}

For this study three types of Sobol Indices are calculated:
\begin{enumerate}
\item \emph{First order Sobol Index:}\\
The first order sensitivity index $S_i$ measures the single effect of the input $h_i$ on the output variance of the model \cite{sobol1993sensitivity,saltelli2010variance}. $S_i$ is defined as
\begin{equation}
S_i=\frac{Var_i}{Var_{Y}}=\frac{Var_i[\mathbb{E}_{\sim i}(Y|h_i)]}{Var_{Y}},
\end{equation}
where the conditional expectation $\mathbb{E}_{\sim i}(Y|h_i)$ denotes the expected value of the output $Y$ when the input $h_i$ is fixed.
\item \emph{Second order Sobol Index:}\\
The second order sensitivity index $S_{ij}$ measures the interaction between $h_i$ and $h_j$ \cite{sobol1993sensitivity,saltelli2010variance}. $S_{ij}$ can be computed as
\begin{equation}
S_{ij}=\frac{Var_{ij}}{Var_{Y}}=\frac{Var_{ij}[\mathbb{E}_{\sim ij}(Y|h_i,h_j)]}{Var_{Y}}-S_i-S_j,
\end{equation} 
where the conditional expectation $\mathbb{E}_{\sim i}(Y|h_i,h_j)$ is the expected value of the output $Y$ when the two input $h_i$ and $h_j$ are fixed.
\item \emph{Total order Sobol Index}:\\
The Total sensitivity index $S_{Ti}$ is called the "Total effect" of a input parameter $h_i$ \cite{nagel2017uncertainty}. This index includes the effect of the first order indices and the effects between the input $h_i$ and all the possible combinations with the other inputs \cite{homma1996importance}. It is defined as
\begin{equation}
S_{Ti}=1- \frac{Var_{\sim i}[\mathbb{E}_i(Y|h_{\sim i})]}{Var_Y},
\end{equation}
where $h_{\sim i}$ refers to all the inputs except $h_i$. For instance, $h_{\sim 1}=h_2,h_3,h_4,h_5,h_6$.
\end{enumerate}

\begin{figure}[H]
\centering\includegraphics[width=1\linewidth]{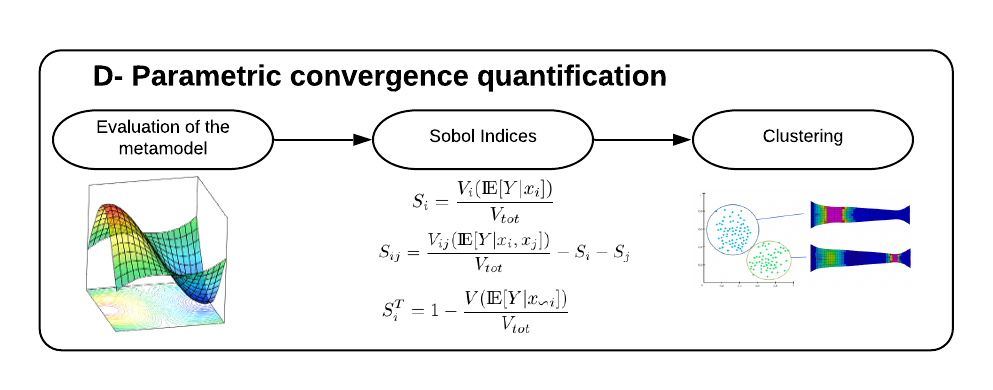}
\caption{Flowchart of the parametric convergence quantification step.}
\label{fig:uncertainty_block}
\end{figure}

For the calculation of the conditional expectation of the Sobol indices  there exist different estimators in order to optimize the number of model evaluations \cite{saltelli2010variance}. For this research, Saltelli algorithm \cite{kucherenko2017different} is implemented to calculate the conditional expectation. In addition, it is important to remark that Sobol Indices are computed for all principal components of $\mathbf{z}^{\star}\in\RR^{k}$. In Fig. \ref{fig:uncertainty_block} it is illustrated the main ideas for the parametric convergence quantification step.

\subsection{Autonomous stopping criteria	}
\label{stopping_criteria}
For the proposed adaptive UQ approach is required a stopping criteria. The variables used to analyze the convergence are both, the Sobol Indices and the percentage of the clusters. These two variables are proposed for the importance in the field of crashworthiness to detect structure modes and main parameters. However, any other sensitive measures can be implemented as a stopping criteria for the reduced space $\mathbf{z^{\star}}$ (e.g. mean, variance, standard deviation, histograms) depending on the problem.

The strategy is based in a comparative approach between levels of different sampling size. The size of the training sets for each level are defined by $\ns^{\ell}=\ncon\ell$ with $\ell=1,2,...,L$. Where $\ncon$ is a constant variable of the number of simulations. This parameter depends on the problem and is defined by the user. For the benchmark problem $\ncon=10$, which leads to obtain the levels:

\begin{itemize}
\item$Level_1\rightarrow\ns^1=10$
\item$Level_2\rightarrow\ns^2=20$
\item$Level_3\rightarrow\ns^3=30$\\
\vdots
\item$Level_L\rightarrow\ns^L=10L$
\end{itemize}

\begin{figure}
\centering\includegraphics[width=0.9\linewidth]{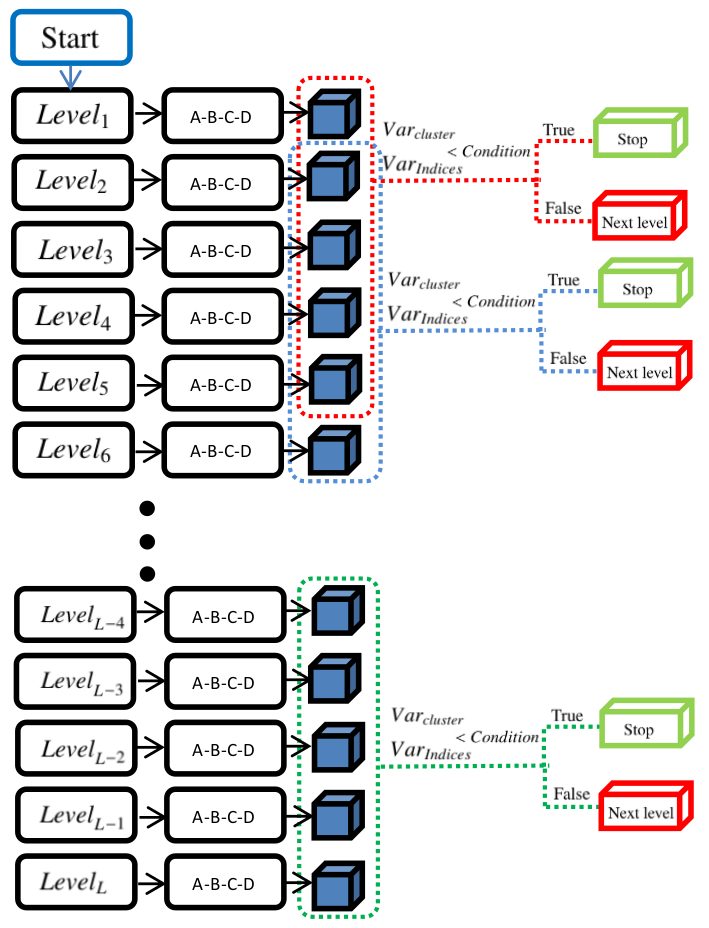}
\caption{Flowchart scheme of the adaptive stopping criteria.}
\label{fig:stop_block}
\end{figure}

Referring to the flowchart in Fig. \ref{fig:stop_block}, the methodology starts with $Level_1$ and the computation of the steps $A,B,C,D$. Then, $Level_2$ is launched with the computation of the steps with a new sampling size $\ns^2$. For the new sampling size of each level it is reused the simulations of the previous level, aiming to evaluate the minimum number of simulations of the full order model. The first levels are launched  until $s$ levels (for the benchmark problem $s=5$). In the last level ($ Level_s $) the variance of the percentage of the modes and the Sobol indices of the previous $ s $ levels are analyzed. If the variances of the stopping variables (Sobol indices and cluster percentage) are sufficiently small with respect to a stopping variance condition, the method stops. If not, the approach goes to a next level with the new increment of sampling size. The stopping variables are always analyzed for the last $s$ levels, meaning that the parameter $s$ defines how stationary is the solution. The process is repeated for each new level until the stability is achieved. In addition, it is necessary to achieve these stopping conditions for each dimension of the reduced space $\mathbf{z}^{\star}\in\RR^{k}$.

\subsection{Uncertainty quantification}
\label{uncertainty_quantification}

Once the methodology is converged, the backward mapping returns $\mathbf{z}^{\star}$ to the corresponding input space $\mathbf{x}$ through $\mathbf{x}^{\star}= \mathcal{G}^{-1}(\mathbf{z^{\star}})$, aiming to develop uncertainty quantification of the input space.

Uncertainty quantification of high-dimensional objects like $\mathbf{x}$ is cumbersome and the outcome is difficult to use as a tool supporting decision making. In that sense, the stochastic assessment focuses in a low-dimensional (even purely scalar) QoI, rather than in a high-dimensional object like $\mathbf{x}$. A specific QoI is introduced as an essential indicator for decision making. 

For this study, the specific QoI is the average plastic strain of the input vector $\mathbf{x}$. However, any QoI function can be analyzed depending of the purpose of the problem. Here, it is represented by a function form $l^{0}(\cdot)$, and for each $\bh^{i}$ and  $\mathbf{x}^{i}$ reads 

\begin{equation}
l^{0}(\mathbf{x}^{i})=\frac{1}{d} \sum_{j=1}^{d}x_j^i . 
\end{equation} 

Statistical measures of $l^{0}(\mathbf{x})$ can be performed (e.g. mean, variance, standard deviation).

\begin{figure}[H]
\centering\includegraphics[width=1\linewidth]{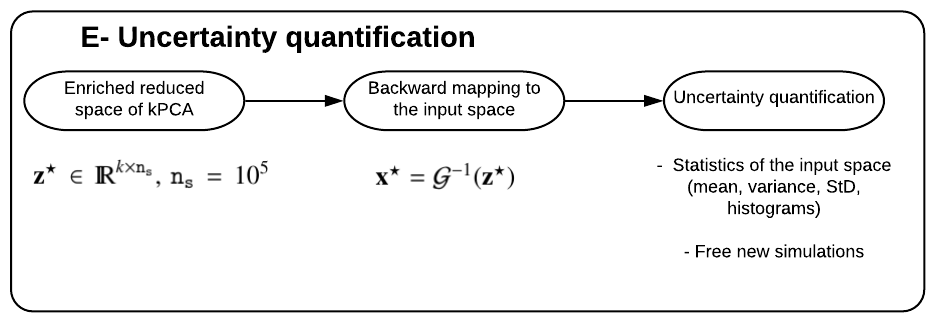}
\caption{Flowchart of the uncertainty quantification step.}
\label{fig:uq_step}
\end{figure}

On the other hand, the fact that the model order reduction strategy is able to recover back the full-order object $\mathbf{x}^{\star}$, is extremely advantageous to represent new simulations with negligible computational cost. Therefore, for any combination of the input parameters $\bh$ it is able reproduce the solution of the vector $\mathbf{x}$ for the physical model. In Fig. \ref{fig:uq_step} it is illustrated the main ideas for the uncertainty quantification step.

\begin{figure*}[htb]
\centering\includegraphics[width=0.8\linewidth]{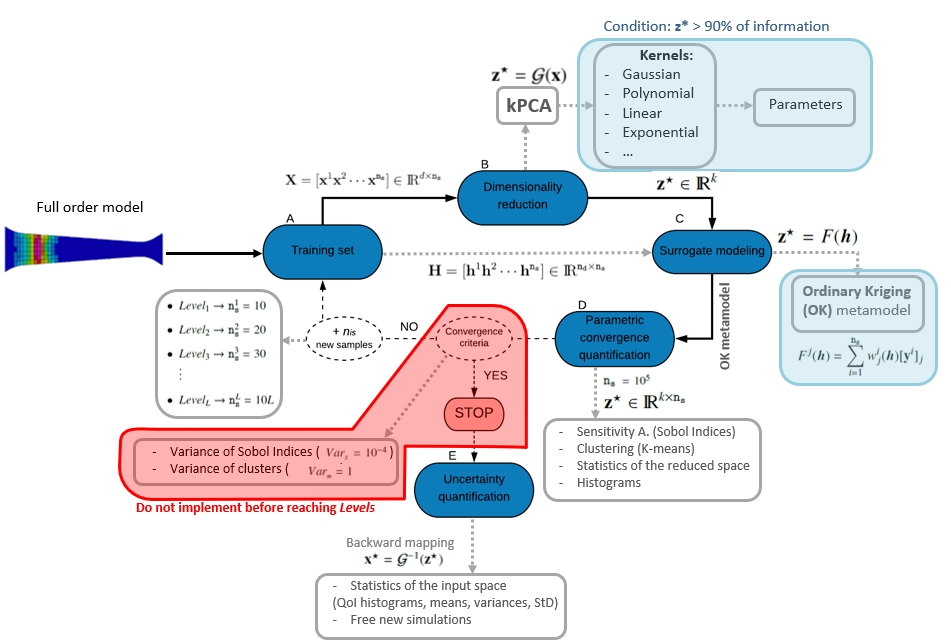}
\caption{Flowchart overview of the proposed adaptive methodology.}
\label{fig:overview}
\end{figure*}

On the other hand, for a better understanding of the previous sections, a more detailed overview of the adaptive method is presented in Fig. \ref{fig:overview}. There are illustrated the 5 most important steps (A, B, C, D, E in navy blue) and the most important information derived from them.

\section{Numerical results}
\label{S:4}
In this section the proposed methodology is implemented for the benchmark problem. The numerical results are divided in two sub-sections: i) The \textit{Reference results}, where the benchmark problem has been evaluated with a vademecum of 3000 simulations, and ii) \textit{UQ adaptive results}, where it is evaluated the methodology and compared with the reference results.
\subsection{Reference results}
\begin{figure}[H]
\centering\includegraphics[width=1\linewidth]{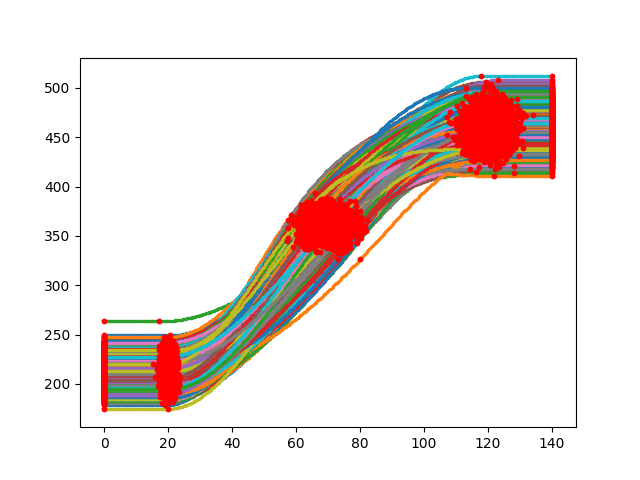}
\caption{Input hardness curves of 3000 samples.}
\label{fig:input_curves}
\end{figure}

The idea of this section is to calculate the reference results, aiming to be compared with the proposed adaptive methodology. For this, the proposed methodology is implemented with a fixed number of samples $\ns$ for the training set. It is considered six stochastic inputs $\bh=[h_1,h_2,h_3,h_4,h_5,h_6]^T$ for the benchmark problem, where each input follows a normal distribution with its corresponding mean and standard deviation described in Table \ref{tab:table1}. For the QoI output it is considered the maximum plastic strain of all the shell elements ($d=329$) from the tapered geometry. For the reference data set, a vademecun of $\ns=3000$ Monte Carlo samples are computed to obtain the output matrix, $\mathbf{X}=[\mathbf{x}^1 \mathbf{x}^2 \cdots\mathbf{x}^{\ns}]\in \RR^{d \times \ns}$. Each vector $\mathbf{x}^i, i=1,2,...,\ns$ store the maximum plastic strain of all the elements for each simulation in the last step of time. In Fig. \ref{fig:input_curves} it is shown the 3000 stochastic hardness curves for each simulation.

The dimensionality reduction problem is implemented with a Polynomial Kernel, $k(\mathbf{x}^i,\mathbf{x}^j)=(<x^i,x^j>+b)^p$ with a coefficient $b=0.1$ and a polynomial degree $p=3$. With this kernel configuration it is reduced the dimension of the problem to the first principal component ($k=1$) with capturing $98,8\%$ of the variance information. Leading to obtain a reduced space $Y=\mathbf{z}^{\star}=[\mathbf{z}^{\star}]_1\in \RR^{k \times \ns}$. Using linear  PCA, 17 principal components ($k=17$) are required to capture the same percentage information, and in consequence 17 metamodels for each dimension. Clearly an improvement is shown with respect to PCA (kPCA $\rightarrow$k=1 and PCA $\rightarrow$k=17). For the metamodel it is implemented OK between the input parameters $\bh$ and the feature space $[\mathbf{z}^{\star}]_1$ with a spherical variogram \cite{oliver2014tutorial}.  For the UQ analysis (statistic measures and Sobol indices) it is evaluated the metamodel with $10^5$ random samples.

\begin{figure}
\centering\includegraphics[width=0.9\linewidth]{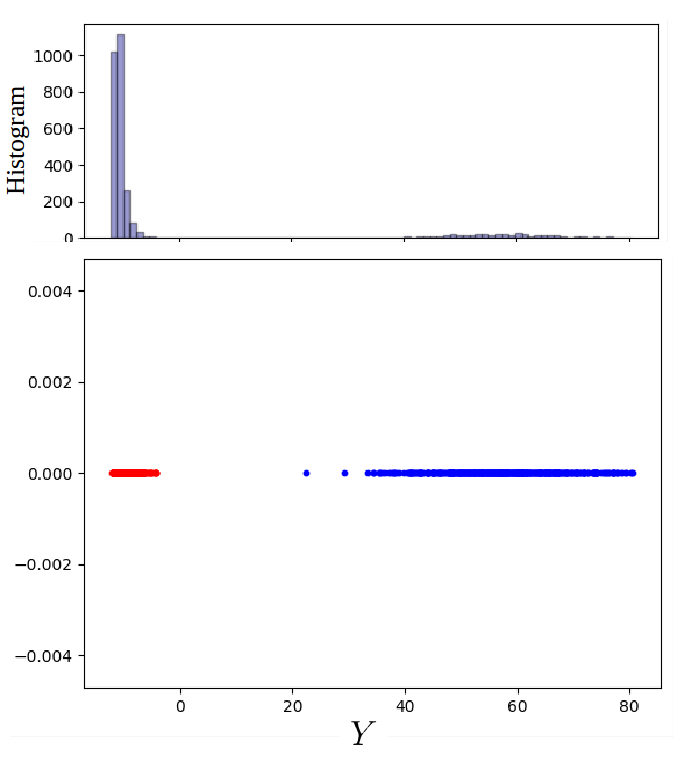}
\caption{Histogram and reduced space $Y$.}
\label{fig:reduced_space}
\end{figure}

In Fig. \ref{fig:reduced_space} it is shown the reduced space of the first principal component $Y=[\mathbf{z}^{\star}]_1$ of kPCA and the corresponding PDF. Clearly, two clusters (modes of the structure) of samples are differentiated. The red and blue samples  are plotted by K-means algorithm \cite{kodinariya2013review}. The percentage of probability for each clusters are:

\begin{itemize}
\item Red cluster: $84.43\%$.
\item Blue cluster: $15.57\%$.
\end{itemize}

\begin{figure}
     \centering
     \subfloat[]{\includegraphics[width=0.4\textwidth]{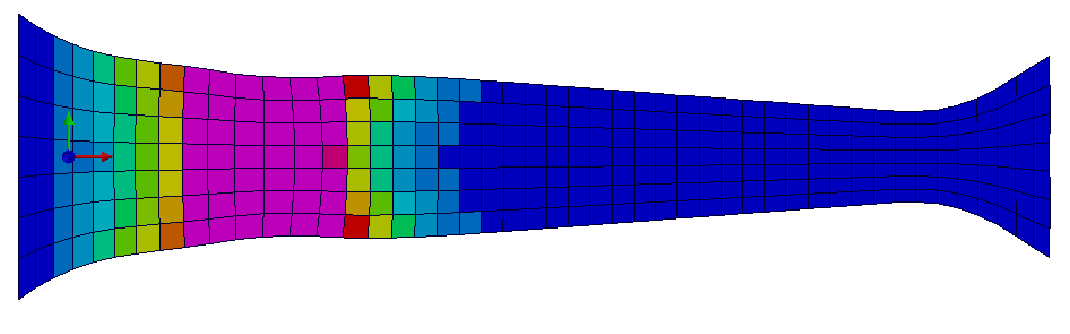}\label{fig:sfig1}}\\
      \subfloat[]{\includegraphics[width=0.4\textwidth]{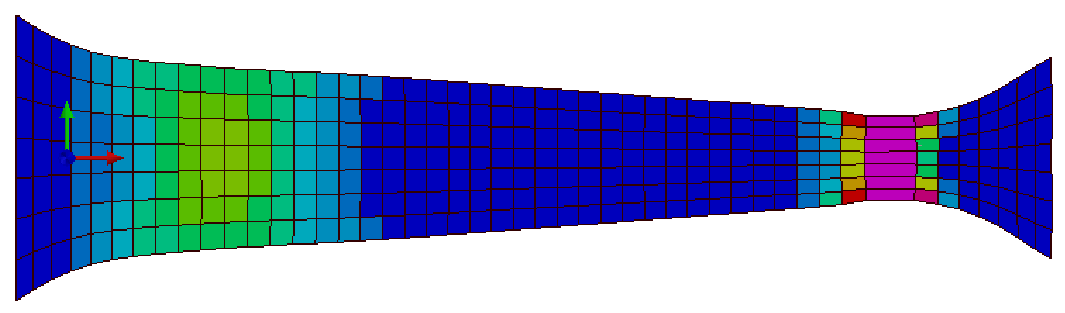}\label{fig:sfig2}}
     \caption{(a) Corresponds to a sample from the red mode of the reduced space $Y$ transformed by backward mapping to the original space of the model. Figure (b) corresponds to a sample from the red mode of the reduced space $Y$ transformed by backward mapping to the original space of the model.}
     \label{fig:modes}
\end{figure}

In Fig. \ref{fig:modes} a solution of the original model for each cluster (red and blue) is shown. It is observed two different behaviours. The structure breaks either on the right (Fig. \ref{fig:sfig1}) or to the left (Fig. \ref{fig:sfig2}).
	
\begin{figure}
\centering\includegraphics[width=1\linewidth]{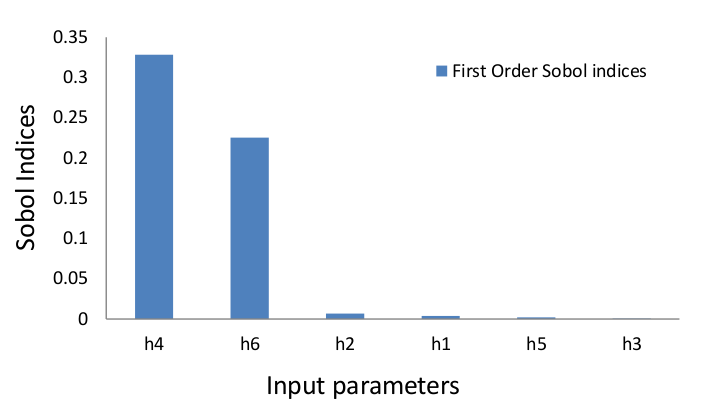}
\caption{First order Sobol indices.}
\label{fig:first_sobol}
\end{figure}

\begin{figure}
\centering\includegraphics[width=1\linewidth]{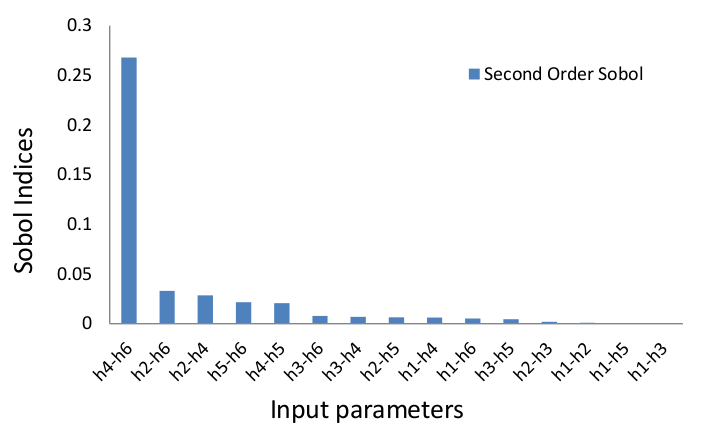}
\caption{Second order Sobol indices.}
\label{fig:second_sobol}
\end{figure}

\begin{figure}
\centering\includegraphics[width=1\linewidth]{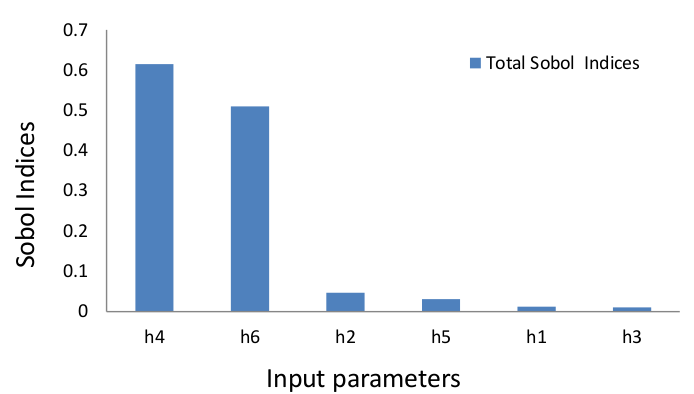}
\caption{Total Sobol indices.}
\label{fig:total_sobol}
\end{figure}

For the Sobol sensitivity analysis it is used Saltelli method to obtain the conditional variance for each index \cite{saltelli2010variance}. In Fig.\ref{fig:first_sobol} it is plotted the values of the first order of Sobol Indices for each input. In Fig. \ref{fig:second_sobol} are shown the second order Sobol indices. This index explains the interaction effect between all the possible pairs of parameters with respect to the output variance output. In Fig. \ref{fig:total_sobol} it is illustrated the 6 Total Sobol indices. This index explains the total effect of an input parameter $h_i$ to the total variance $Var_{Y}$.This measures the effect of the output variance of $h_i$, with respect to any variable and any order of interaction between parameters. 

From the above presented results, in Table \ref{tab:caption} it is illustrated the most important variables to take into account as a reference values for the next section \ref{adaptive} where the adaptive methodology is tested.\\

\begin{table}[ht]
\centering
\caption{Reference values from a vademecum of 3000 training samples.}
\begin{tabular}{llrll}
\toprule
\textbf{ Clustering modes  } &      \\
\toprule
\midrule
Left mode = 84.43\% \\
Right mode = 15.57\%   \\

\toprule
\textbf{Sobol Indices}   &      \\
\toprule
\midrule
\multicolumn{2}{c}{\emph{First order}} \\
\hline
$S_4$  = 0.35 \\
$S_6$ = 0.23  \\

\toprule
\multicolumn{2}{c}{\emph{Second order}} \\
\hline
$S_{46}$  =  0.26 \\
\toprule

\multicolumn{2}{c}{\emph{Total indices}} \\
\hline
$S_{T4}$  = 0.62 \\
$S_{T6}$ = 0.51   \\
\bottomrule
\end{tabular}
\label{tab:caption}
\end{table}%

\subsection{UQ adaptive results}
\label{adaptive}
In this section it is presented the numerical results to validate the performance of the adaptive methodology applied to the benchmark problem described in Section \ref{S:2}. The approach is implemented with a polynomial kernel (degree $p=3$) for the dimensionality reduction problem and OK for the surrogate modeling from $\bh$ to $Y$. For the benchmark, $Y$ coincides with the first component of the reduced space using kPCA. The criterion to stop the adaptive methodology is based in a variance criterion. The approach stops once the variance of the previous 5 levels ($s=5$) achieves the order of $Var_{_{S}}=10^{-4}$ (for the Sobol Indices) and $Var_{_{m}}=1$ (for the mode percentage).

\begin{figure}
\centering\includegraphics[width=0.65\linewidth]{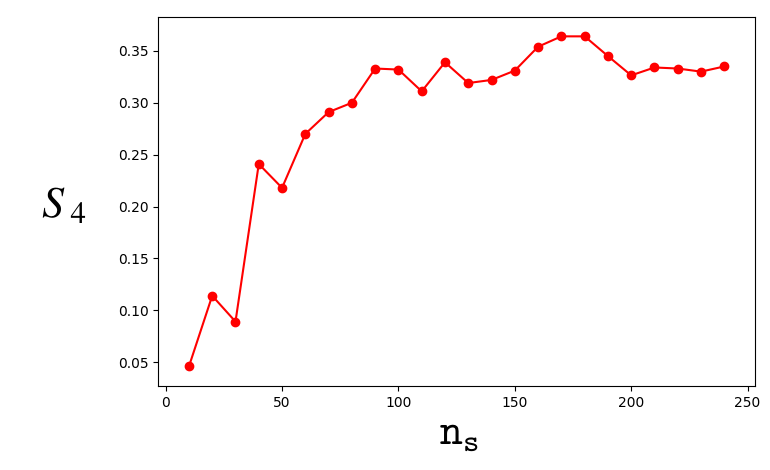}\\
\centering\includegraphics[width=0.65\linewidth]{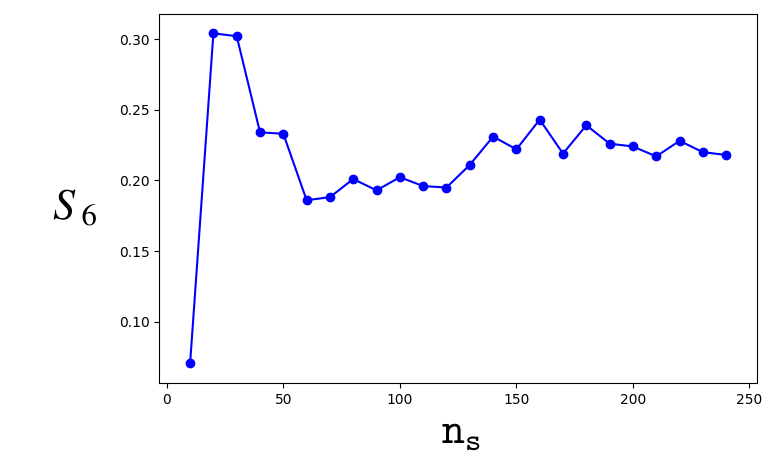}\\
\centering\includegraphics[width=0.65\linewidth]{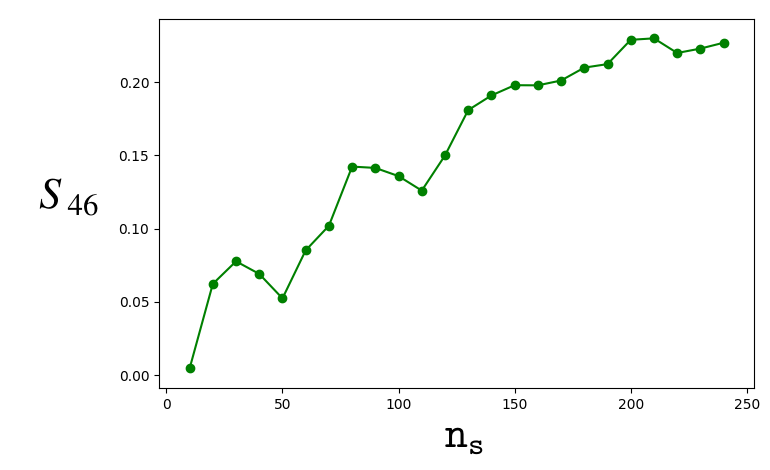}\\
\centering\includegraphics[width=0.65\linewidth]{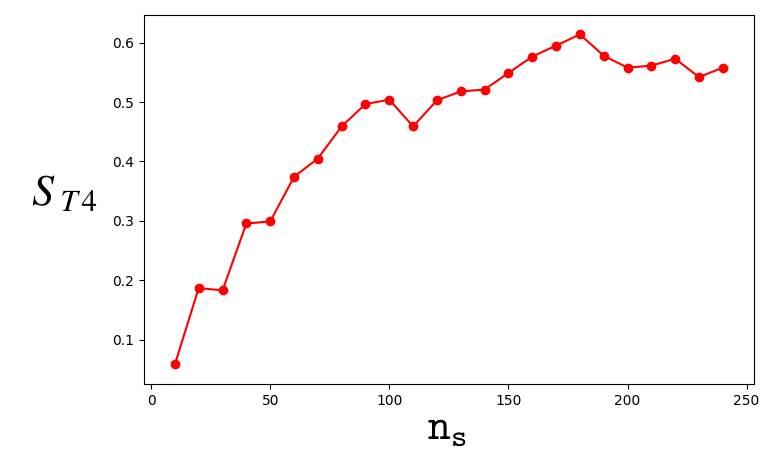}\\
\centering\includegraphics[width=0.65\linewidth]{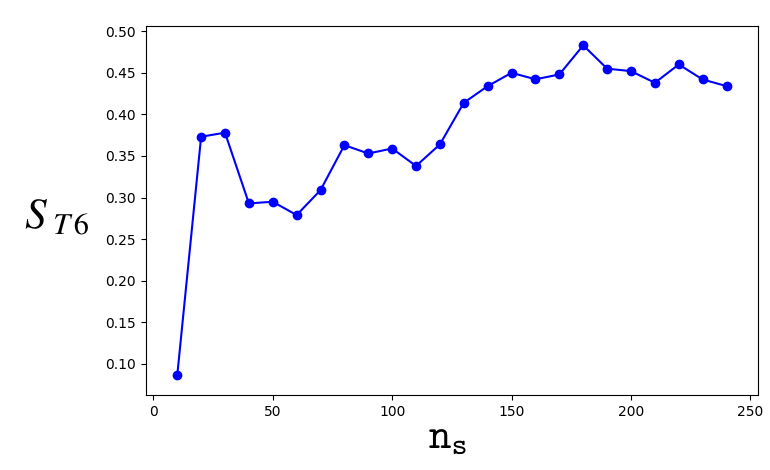}\\
\centering\includegraphics[width=0.65\linewidth]{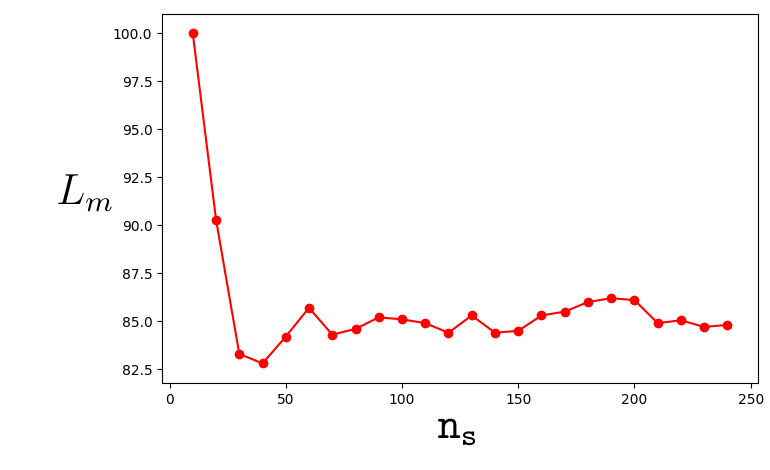}
\caption{Evolution plots of the First Sobol Indices ($S_4$, $S_6$), Second order Sobol Indice ($S_{46}$), Total Sobol Indices ($S_{T4}$, $S_{T6}$) and the percentage of the left mode ($L_m$).}
\label{fig:hierarchy_results}
\end{figure}

In Fig. \ref{fig:hierarchy_results} it is plotted the evolution of the Sobol indices and the cluster percentage for each level of sampling size.  We can analyze that the last 5 sample points of each graph have small variability, corroborating that the stopping criteria is accomplished. The method stops in $Level_{24}$ ($L=24$) with a training set of $\ns^{24}=240$. 

In Table \ref{tab:comparative} it is compared the results obtained with the training set of 3000 samples (\textit{Vade.}) with respect to the adaptive approach with 240 training samples (\textit{Adapt.}).

\begin{table}[ht]
\centering
\caption{Comparison results between reference vademecum (3000 training samples) with respect to the adaptive methodology (240 training samples).}
\begin{tabular}{cccccccc}
\toprule
 &$L_m$&$R_m$&$S_4$&$S_6$&$S_{46}$&$S_{T4}$&$S_{T6}$ \\  
\toprule
\midrule		
\textit{Vade.} &  $84.43\%$&$15.57\%$ &0.35 &0.23 &0.26 &0.62 &0.51\\
\midrule
\textit{Adapt.}&$84.81\%$&$15.19\%$&0.34&0.22&0.23&0.56&0.43\\
\bottomrule
\end{tabular}
\label{tab:comparative}
\end{table}%

\begin{figure}[H]
\centering\includegraphics[width=0.9\linewidth]{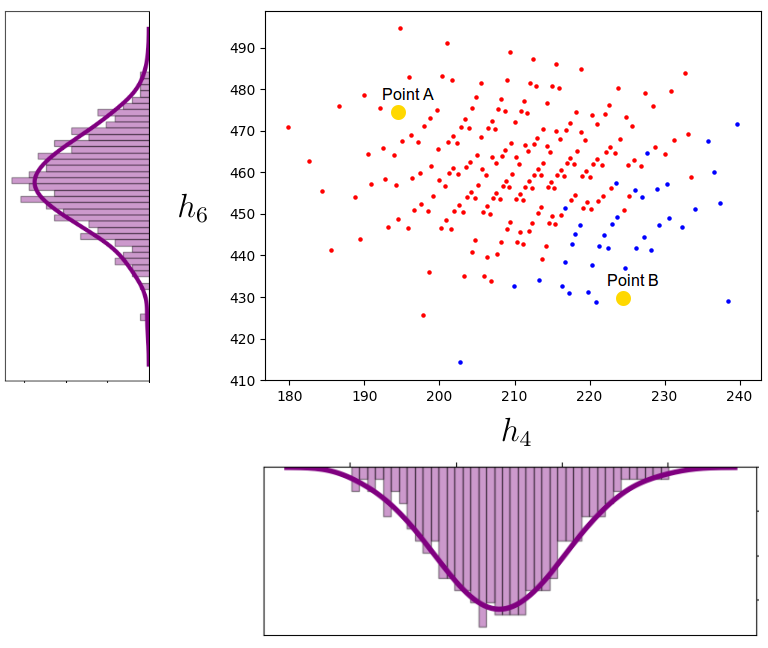}
\caption{Scatter plot between the inputs $h_4$ and $h_6$. Red samples corresponds to the left mode and blue samples to the right mode. Point A ($h_1=22$, $h_2=60$, $h_3=128$, $h_4=195$, $h_5=333$, $h_6=472$). Point B ($h_1=18.5$, $h_2=65$, $h_3=122$, $h_4=224.5$, $h_5=365$, $h_6=430$).}
\label{fig:scatter}
\end{figure}

The surrogate model constructed with 240 training samples brings a powerful tool. Statistical measures (mean, variance and standard deviation) and scattered plots offer and interesting analysis to understand an analyze the cause of each structure mode. Fig. \ref{fig:scatter} shows the scatter plot between the two main parameters $h_4$ and $h_6$. The sample points are coloured in red or in blue, depending on the structure mode (left=red, right=blue).

Here it is clearly visible two differentiated areas of color points. This means that the relation between the parameters $h_ 4$ and $h_ 6$ defines practically the whole behaviour of each mode.

\begin{figure}[H]
\subfloat[VPS/Pamcrash result (Point A).]{
\centering\includegraphics[width=0.49\linewidth]{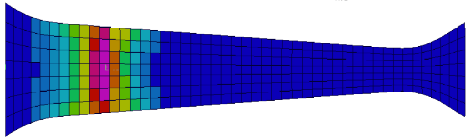}}
\subfloat[VPS/Pamcrash result (Point B).]{
\centering\includegraphics[width=0.46\linewidth]{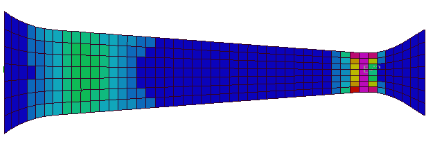}}\\

\subfloat[Backward with PCA (Point A).]{
\centering\includegraphics[width=0.49\linewidth]{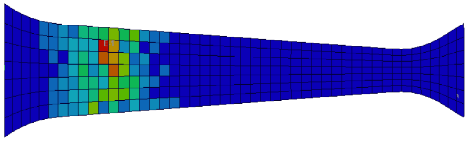}}
\subfloat[Backward with PCA (Point B).]{
\centering\includegraphics[width=0.48\linewidth]{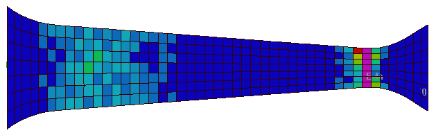}}\\

\subfloat[Backward with kPCA (Point A).]{
\centering\includegraphics[width=0.49\linewidth]{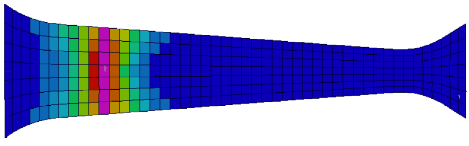}}
\subfloat[Backward with kPCA (Point B).]{
\centering\includegraphics[width=0.49\linewidth]{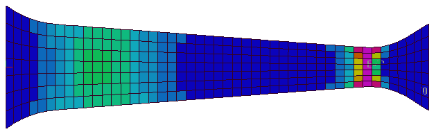}}\\

\caption{(a) and (b) illustrates the full order simulation with VPS/Pamcrash for the points A and B. Also (c), (d), (e) and (f) show the backwards from $Y$ to the original space $\mathbf{x}$ with PCA and kPCA.}
\label{fig:backward}
\end{figure}

Each training sample needs considerable CPU resources and time.  This method allows to obtain new simulations with a negligible consumption of time. In the scatter plot (Fig.\ref{fig:scatter}) a new point $\bh$ for each color area is selected in a empty zone of the plot (yellow points) to backward to the original space $\mathbf{x}$. In Fig. \ref{fig:backward} it is compared the full order computational samples with respect to the corresponding backward samples from $Y$ to $\mathbf{x}$ with PCA and kPCA techniques. The performance of PCA and kPCA is shown to illustrate the improvement of kPCA for this crashworhtiness model.

kPCA clearly shows better performance in the backward mapping. Taking into account that with kPCA the first principal component $[\mathbf{z}^{\star}]_1$ contains $98.8\%$ of information and PCA only $77.7\%$.

\begin{figure}[H]
\centering\includegraphics[width=1\linewidth]{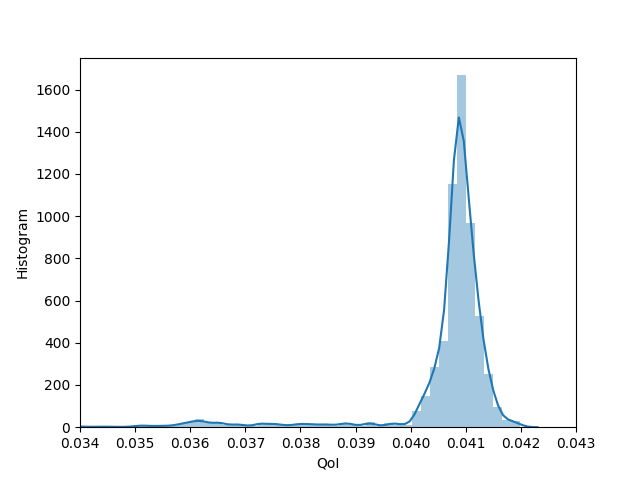}
\caption{Histogram of the QoI with $\ns=10^5$.}
\label{fig:QoI_histogram}
\end{figure}

On the other hand, in Fig. \ref{fig:QoI_histogram} it is illustrated the histogram of the specific QoI function $l^0(\mathbf{x})$. Depending on the problem, the QoI is sufficient i some cases, for decision making \cite{2102.07673}, since the QoI summarizes the information contained in $\mathbf{x}$.  Here, the QoI corresponds to the average of vector $\mathbf{x}$. The statistical measures of the QoI are: mean=$0.0407$, variance=$1.12e-06$ and standard deviation=$0.0011$. The histogram presents a normal distribution centred approximately at $0.041$ and a long tail on the left. The samples falling in the normal distribution corresponds to the left structure mode. Otherwise, the samples that fall into the distribution tail makes reference to the samples with the structure mode on the right. 

\section{Discussion}
\label{S:5}

In this article, the proposed UQ methodology combining kPCA and surrogate modeling for nonlinear problems is implemented for a real industrial crash problem. \MRAText{The uncertainty of the problem is characterized with 6 input random parameters defining the hardness curve of the material model. The maximum plastic strain for all the elements in the last time step is considered as QoI of the model.}

The methodology is implemented with a polynomial kernel. The convergence is achieved with 240 samples for the training set with a stopping criteria of a variance condition of $Var_{_{S}}=10^{-4}$ (for the Sobol Indices) and $Var_{_{m}}=1$ (for the mode percentage). The method detected two structure modes (clusters). The big mode approximately with $84\%$ and a small mode with $16\%$ of probability. In the biggest structure mode, the pyshical model concentrates high values of plastic strain in the left part. In contrast, in the small mode they are on the right area. Moreover, the main influence parameters for the output are $h_4,h_6$ ($h_4\rightarrow 34\%$ and $h_6 \rightarrow 22\%$) for the first order, and $h_4-h_6$ ($h_4-h_6 \rightarrow 23\%$) for the second order of sensitivity. This means that the relation between these two parameters is defining practically all the behaviour of the model. In addition, the total Sobol Indices $S_{T4}$ and $S_{T6}$ are showing similar relation with respect to the first and second order of Sobol Indices. This emphasizes that the other parameters $h_1,h_2,h_3,h_5$ have low influence to the output. Also, the specific QoI function $l^0(\cdot)$ shows a normal distribution with a long tail in the left allowing to facilitate the understanding of the high dimension vector $\mathbf{x}$.

\section{Conclusions}
\label{S:6}

Uncertainty quantification in crash simulation is a highly demanding research field for the automotive industry. Its nonlinear behaviours combined with hidden structure modes leads to a challenging task for UQ analysis. High dimensional  outputs for the quantity of interest can be a challenging problem for surrogate modelling by suffering the curse of dimensionality. \MRAText{This problem is relevant for automotive engineering and, despite the fact that only six input parameters are assumed to have stochastic nature (6 input dimensions are not awakening the curse of dimensionality), the dimensionality reduction is still pertinent to simplify the output of interest to be analyzed.}

Additionally, each evaluation of a full vehicle model needs around hours. Therefore, classic approaches as Monte Carlo are not viable. \MRAText{Here, a benchmark model requiring a significantly lower computational time, and of special interest for automotive industry,  is used to demonstrate the performance of the adaptive methodology with a data driven strategy.}

This paper presents an adaptive methodology for crashworthiness combining dimensionality reduction and surrogate modelling for multi-purpose UQ analysis. The methodology evaluates limited set of samples of the high order model guaranteeing a good precision. The  problem of dimensionality reduction for the outputs is tackle using kPCA in such a way OK for metamodelling  the reduced space of kPCA and the input samples $\bh^i,i=1,2,\cdots,\ns$. Moreover, key information as cluster detection, percentage of clusters, sensitivity analysis, statistics and free new simulations provide robust and reliable information supporting decision making. Having access to these descriptive behaviours of the model is a great advantage for CAE departments. Overall, the proposed methodology requires significantly less memory and resources compared to classic methods as Monte Carlo for calculating UQ and sensitivity indices for the high dimensional outputs.

The results from the industrial benchmark verified the performance and accuracy of the proposed methodology with respect to a vademecum approach of 3000 samples for the training set. The method can be extended and applied for other disciplines (e.g. aerodynamics, occupant safety, aeroacoustic, among others) with uncertainty inputs, nonlinear responses and high dimensional outputs. The methodology is presented with kPCA and OK for dimensionality reduction and surrogate modeling, respectively. Nevertheless, other dimensionality reduction techniques as Isometric Mapping or Locally Embeddings can be implemented as well. In addition, other metamodel techniques (regression, interpolation) can be used as well depending on the data. 

\MRAText{The combination of dimensionality reduction and surrogate models with an adaptive approach for multi-purpose information produces accurate solutions with an affordable computational cost, accounting also for the uncertainty, that is assessing the credibility of the simulation.} Particularly in the context of crashworthiness UQ, the computational cost is a key issue and a driving force for the research developments in the field. Since, increasing accuracy requires a higher computational effort, finding a trade-off between these two factors is a critical concern for last decision making. This paper intends to provide tools to achieve accurate and credible crashworthiness industrial simulations at an acceptable computational effort.

\section*{Acknowledgements}
This work is partially funded by Generalitat de Catalunya (Grant Number 1278 SGR 2017-2019 and Pla de Doctorats Industrials 2017 DI 058) and Ministerio de Econom\'ia y Empresa and Ministerio de Ciencia, Innovaci\'on y Universidades (Grant Number DPI2017-85139-C2-2-R).

\section*{Compliance with Ethical Standards}

The authors declare that they have no conflict of interest.

\bibliographystyle{plain}      
\bibliography{Paperbib}   







\end{document}